\documentclass[english,english,pra,english,preprint,amsmath,amssymb,aps,longbibliography,showkeys, titlepage]{revtex4-2}
\usepackage{lmodern}
\usepackage[T1]{fontenc}
\usepackage[latin9]{inputenc}
\usepackage{amsmath}
\usepackage{amsthm}
\usepackage{amssymb}
\usepackage{graphicx}

\makeatletter
\usepackage[T1]{fontenc}
\usepackage{amsmath}
\usepackage{amsthm}
\usepackage{amssymb}
\usepackage{graphicx}
\usepackage{tikz-feynman}
\tikzfeynmanset{compat=1.1.0}

\usepackage{graphicx}
\usepackage{float}
\usepackage{xcolor}
\usepackage[hypertexnames=false]{hyperref}
\hypersetup{
	breaklinks = true,
    colorlinks = true,
    citecolor = {blue},
	urlcolor = {blue},
	linkcolor = {blue}
}

\makeatother

\usepackage{babel}
\begin{document}
\title{Quantum Inspiration, Classical Advantage: Dequantized particle algorithm
for the nonlinear Vlasov-Poisson system}
\author{Hong Qin}
\email{hongqin@princeton.edu }

\affiliation{Princeton Plasma Physics Laboratory, Princeton University, Princeton,
NJ 08540}
\affiliation{Department of Astrophysical Sciences, Princeton University, Princeton,
NJ 08540}
\author{Michael Q. May}
\email{mqmay@princeton.edu }

\affiliation{Princeton Plasma Physics Laboratory, Princeton University, Princeton,
NJ 08540}
\affiliation{Department of Astrophysical Sciences, Princeton University, Princeton,
NJ 08540}
\author{Jacob Molina}
\email{jmmolina@princeton.edu }

\affiliation{Princeton Plasma Physics Laboratory, Princeton University, Princeton,
NJ 08540}
\affiliation{Department of Astrophysical Sciences, Princeton University, Princeton,
NJ 08540}
\begin{abstract}
We present a dequantization algorithm for the Vlasov--Poisson (VP)
system, termed the dequantized particle algorithm, by systematically
dequantizing the underlying many-body quantum theory. Starting from
the second-quantized Hamiltonian description, we derive a finite-dimensional
dequantized system and show that it furnishes a structure-preserving
discretization of the Schrödinger--Poisson (SP) equations. Through
the Wigner or Husimi transformations, this discretization provides
an efficient approximation of the VP system when quantum effects are
negligible. Unlike conventional structure-preserving algorithms formulated
in 6D phase space, this dequantized particle algorithm operates in
3D configuration space, potentially offering more compact and efficient
representations of physical information under appropriate conditions.
A numerical example of the classical nonlinear two-stream instability,
simulated using merely 97 dequantized particles, demonstrates the
efficiency, accuracy, and conservation properties of the algorithm
and confirms its potential as a foundation for developing quantum
and quantum-inspired classical algorithms for kinetic plasma dynamics. 
\end{abstract}
\maketitle
The dynamics of a collection of $N$ identical bosonic charged particles
interacting through Coulomb potential can be described by the Hamiltonian
\begin{equation}
H=\sum_{i=1}^{N}\frac{\boldsymbol{p}_{i}^{2}}{2m}+\frac{1}{2}\sum_{i,j=1}^{N}\frac{q^{2}}{\left|\boldsymbol{x}_{i}-\boldsymbol{x}_{j}\right|}\,,
\end{equation}
where $q$ and $m$ are the charge and mass of the particles. In quantum
many-body theory, $\boldsymbol{p}_{i}=i\hbar\partial/\partial\boldsymbol{x}_{i}$
and the system is governed by the Schrödinger equation 
\begin{equation}
\frac{\partial}{\partial t}\psi(\boldsymbol{x}_{1},...,\boldsymbol{x}_{N},t)=\frac{1}{i\hbar}H\psi(\boldsymbol{x}_{1},...,\boldsymbol{x}_{N},t)\,.\label{eq:MSchro}
\end{equation}
In classical theory, $\boldsymbol{p}_{i}$ $(i=1,...,N)$ are independent
variables of the $N$-particle phase space $\boldsymbol{z}_{i}=(\boldsymbol{x}_{i},\boldsymbol{p}_{i})$
$(i=1,...,N)$, and the dynamics are governed by Hamilton's equation
\begin{equation}
\frac{d\boldsymbol{z}_{i}}{dt}=\left\{ \boldsymbol{z}_{i},H\right\} .\label{eq:HamEq}
\end{equation}
The many-body quantum theory recovers the classical theory when the
characteristic kinetic energy of the particles is much larger than
the characteristic Coulomb potential, $\boldsymbol{p}_{i}^{2}/2m\gg q^{2}/\left|\boldsymbol{x}_{i}-\boldsymbol{x}_{j}\right|$.
To simplify notation, we will overload the symbol $H$ to denote both
quantum operators and classical Hamiltonian functions, depending on
the context.

In classical theory, when the correlation between particles is negligible,
the system can be described by a one-particle distribution function
$f(\boldsymbol{x},\boldsymbol{v},t)$, which satisfies the Vlasov-Poisson
(VP) equations,

\begin{align}
\frac{\partial f}{\partial t} & +\boldsymbol{v}\cdot\frac{\partial f}{\partial\boldsymbol{x}}-\frac{q}{m}\boldsymbol{\nabla}\phi\cdot\frac{\partial f}{\partial\boldsymbol{v}}=0\,,\label{eq:Vla}\\
\nabla^{2}\phi & =-4\pi q\left(\int d^{3}vf-\frac{N}{L^{3}}\right)\,,\label{eq:Poi}
\end{align}
where $\boldsymbol{v}\equiv\boldsymbol{p}/m$, $L$ is the linear
size of the system, and a uniform background charge density is assumed
such that the system is overall charge neutral.

The VP system has many important applications in plasma physics, fusion
energy, and stellar dynamics. Numerical algorithms for the VP equations---and
the more general Vlasov--Maxwell (VM) equations---have long been
an active area of research in these fields. In particular, recent
years have witnessed significant progress in the development of classical
structure-preserving algorithms for the VP and VM systems \citep{Cary93,squire2012geometric,Xiao2013,xiao2015explicit,xiao2015variational,he2015Hamiltonian,he2016hamiltonian,qin2016canonical,kraus2017gempic,burby2017finite,xiao2017local,Xiao2018review,Xiao2019field,Xiao2021Explicit,Glasser2020,Wang2021,Kormann2021,Perse2021,Glasser2022,CamposPinto2022,Burby2023},
including quantum-inspired matrix product state (MPS) method \citep{Ye2022},
quantum algorithms for the linear VP and VM systems \citep{Engel2019,Novikau2022,Ameri2023,Toyoizumi2024,Miyamoto2024}
and the nonlinear VP system \citep{May2025}, and quantum-classical
hybrid algorithm for the nonlinear VM system \citep{Higuchi2024}.
However, quantum algorithms currently available for these systems
are not yet practical on present-day quantum hardware, especially
when compared to state-of-the-art structure-preserving algorithms
running on classical computers. Nevertheless, it is anticipated that
quantum simulations will eventually demonstrate advantages for certain
classes of problems on future quantum computers. Moreover, studying
quantum algorithms may also lead to efficient quantum-inspired classical
algorithms, which are referred to as dequantization algorithms in
this context.

In the present study, we develop such a dequantization algorithm for
the nonlinear VP system, termed the dequantized particle algorithm,
by dequantizing the many-body quantum theory specified by Eq.\,(\ref{eq:MSchro}).
This procedure yields a finite-dimensional, structure-preserving discretization
of the Schrödinger--Poisson (SP) equations, whose solutions approximate
those of the VP system when quantum effects are negligible. Compared
with established structure-preserving algorithms for the VP system
in 6D phase space, the dequantized particle algorithm offers an alternative
formulation in 3D configuration space. Moreover, it provides a two-way
bridge that can also be used to develop quantum algorithms for the
VP system.

We start from the quantum many-body system governed by Eq.\,(\ref{eq:MSchro}).
The Hamiltonian of the system can be equivalently cast into the following
form using the technique of second quantization \citep{Fetter2012,Mahan2000},

\begin{align}
H & =\sum_{ij}H_{ij}\hat{a}_{i}^{\dagger}\hat{a}_{j}+\frac{1}{2}\sum_{ijln}\hat{a}_{i}^{\dagger}\hat{a}_{j}^{\dagger}V_{ijln}\hat{a}_{l}\hat{a}_{n},\label{eq:H-SMB}\\
H_{ij} & \equiv-\int\psi_{i}^{*}(\boldsymbol{x})\frac{\hbar^{2}}{2m}\nabla^{2}\psi_{j}(\boldsymbol{x})d^{3}\boldsymbol{x},\\
V_{ijln} & \equiv\int\psi_{i}^{*}(\boldsymbol{x}_{1})\psi_{j}(\boldsymbol{x}_{1})\frac{q^{2}}{\left|\boldsymbol{x}_{1}-\boldsymbol{x}_{2}\right|}\psi_{l}^{*}(\boldsymbol{x}_{2})\psi_{n}(\boldsymbol{x}_{2})\,d^{3}\boldsymbol{x}_{1}d^{3}\boldsymbol{x}_{2}.
\end{align}
Here, $\psi_{n}(\boldsymbol{x})\,\,(n\in\mathbb{N})$ is a set of
suitable basis functions on one-particle configuration space, and
$\hat{a}_{n}^{\dagger}$ and $\hat{a}_{n}\,\,(n\in\mathbb{N})$ are
the corresponding creation and annihilation operators. Since the free
particle Hamiltonian is $\hbar^{2}\nabla^{2}/2m$, we will take $\psi_{n}(\boldsymbol{x})=\exp(-i\boldsymbol{k}_{n}\cdot\boldsymbol{x})L^{-3/2}$
with $\boldsymbol{k}_{n}=2\pi n/L$. Note that $n$ is a vector index.
For this choice of basis functions,

\begin{align}
H & =\sum_{l}\frac{k_{l}^{2}\hbar^{2}}{2m}\hat{a}_{l}^{\dag}\hat{a}_{l}+\sum_{l,n,g\neq0}\frac{4\pi q^{2}}{L^{3}k_{g}^{2}}\hat{a}_{l-g}^{\dag}\hat{a}_{n+g}^{\dag}\hat{a}_{l}\hat{a}_{n}\thinspace,\label{eq:Haa}
\end{align}
where use has been made of 
\begin{equation}
V(\boldsymbol{k})\equiv\int\frac{q^{2}}{\left|\boldsymbol{x}\right|}\exp(-i\boldsymbol{k}\cdot\boldsymbol{x})d^{3}\boldsymbol{x}=\frac{4\pi q^{2}}{k^{2}}.
\end{equation}
The dynamics of the system is governed by 
\begin{equation}
\dot{\hat{a}}_{j}=\frac{1}{i\hbar}[\hat{a}_{j},H]=\frac{-i\hbar k_{j}^{2}}{2m}\hat{a}_{j}+\frac{2\pi q^{2}}{i\hbar L^{3}}\sum_{l,g\neq0}\frac{1}{k_{g}^{2}}\hat{a}_{l}^{\dagger}\hat{a}_{l-g}\hat{a}_{j+g}\thinspace.\label{eq:adot-1}
\end{equation}

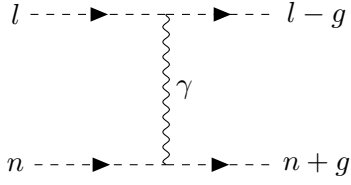
\begin{figure}
\begin{tikzpicture}
  \begin{feynman}
	\vertex (a) at (0, 1) {$l$};
      \vertex (b) at (0, -1) {$n$};
      \vertex (c) at (2, 1);
      \vertex (d) at (2, -1);
      \vertex (e) at (4, 1) {$l - g$};
      \vertex (f) at (4, -1) {$n + g$};
      \diagram* {
        (a) -- [dashed, with arrow=0.5] (c) -- [dashed, with arrow=0.5] (e),
        (b) -- [dashed, with arrow=0.5] (d) -- [dashed, with arrow=0.5] (f),
        (c) -- [photon, edge label=$\gamma$] (d),
      };
  \end{feynman}
\end{tikzpicture}

\caption{Two interacting particles exchange momentum via a photon that carries
momentum $\hbar\boldsymbol{k}_{g}$.}
\label{Fig:FD}
\end{figure}

The first term on the right hand side of Eq.\,(\ref{eq:Haa}) is
the free-particle operator, and the second term represents the interaction
between two particles through Coulomb interaction, as illustrated
in Fig.\,\ref{Fig:FD}.

For practical calculations, it is necessary to truncate the system
for a finite number of $\hat{a}_{j}$. However, it is essential that
the truncated system preserves all the important structures of the
original system, including Hermiticity and the conservation of particles,
energy, and momentum. To this end, we select a finite set $J$ of
vector indices and discretize (truncate) the system as follows, 
\begin{align}
H_{d} & \equiv\sum_{l\in J}\frac{k_{l}^{2}\hbar^{2}}{2m}\hat{a}_{l}^{\dag}\hat{a}_{l}+\sum_{l,n,l-g,n+g\in J}^{g\neq0}\frac{4\pi q^{2}}{L^{3}k_{g}^{2}}\hat{a}_{l-g}^{\dag}\hat{a}_{n+g}^{\dag}\hat{a}_{l}\hat{a}_{n}\thinspace,\label{eq:Hdaa}
\end{align}

\begin{equation}
\dot{\hat{a}}_{j}=\frac{1}{i\hbar}[\hat{a}_{j},H_{d}]=\frac{-k_{j}^{2}i\hbar}{2m}\hat{a}_{j}+\frac{2\pi q^{2}}{i\hbar L^{3}}\sum_{l,l-g,j+g\in J}\frac{1}{k_{g}^{2}}\hat{a}_{l}^{\dagger}\hat{a}_{l-g}\hat{a}_{j+g}\thinspace,\,\,\,\,j\in J.\label{eq:adot}
\end{equation}
It is crucial to note that we require $l,n,l-g,n+g\in J,$ but $g\in J$
is not guaranteed or required. Equations (\ref{eq:Hdaa}) and (\ref{eq:adot})
include the term $k_{g}^{2}$, for which the corresponding $\hat{a}_{g}$
may not exist in the discretized system $H_{d}.$ Physically, $\hbar\boldsymbol{k}_{g}$
is momentum of the gauge photon mediating the interaction between
two particles, see Fig.\,\ref{Fig:FD}.

It is straightforward to see that because of this specific truncation
method, $H_{d}$ remains Hermitian after the truncation. Obviously,
$H_{d}$ is conserved for the discretized system, i.e., $\left[H_{d},H_{d}\right]=0.$
From a physics perspective, the discrete system also conserves the
total number of particles, $N_{d}\equiv\sum_{l\in J}\hat{a}_{l}^{\dag}\hat{a}_{l}$,
because in the expression of $H_{d}$, the number of creation operators
exactly balances the number of annihilation operators. Mathematically,
we have $\left[H_{d},N_{d}\right]=0,$ which can be proved by applying
the Leibniz rule as follows. For every quartet $l,m,l-g,n+g\in J$,
\begin{align}
\left[\hat{a}_{l-g}^{\dag}\hat{a}_{n+g}^{\dag}\hat{a}_{l}\hat{a}_{n},\sum_{j\in J}\hat{a}_{j}^{\dag}\hat{a}_{j}\right] & =\left[\hat{a}_{l-g}^{\dag},\sum_{j\in J}\hat{a}_{j}^{\dag}\hat{a}_{j}\right]\hat{a}_{n+g}^{\dag}\hat{a}_{l}\hat{a}_{n}+\hat{a}_{l-g}^{\dag}\left[\hat{a}_{n+g}^{\dag},\sum_{j\in J}\hat{a}_{j}^{\dag}\hat{a}_{j}\right]\hat{a}_{l}\hat{a}_{n}\nonumber \\
 & +\hat{a}_{l-g}^{\dag}\hat{a}_{n+g}^{\dag}\left[\hat{a}_{l},\sum_{j\in J}\hat{a}_{j}^{\dag}\hat{a}_{j}\right]\hat{a}_{n}+\hat{a}_{l-g}^{\dag}\hat{a}_{n+g}^{\dag}\hat{a}_{l}\left[\hat{a}_{n},\sum_{j\in J}\hat{a}_{j}^{\dag}\hat{a}_{j}\right]\nonumber \\
 & =(-1-1+1+1)\hat{a}_{l-g}^{\dag}\hat{a}_{n+g}^{\dag}\hat{a}_{l}\hat{a}_{n}=0\,.\label{eq:aaaa}
\end{align}
Similarly, the momentum operator, $P_{d}\equiv\sum_{n\in J}k_{n}\hat{a}_{n}^{\dag}\hat{a}_{n},$
of the discrete system is conserved because for each term $\hat{a}_{l-g}^{\dag}\hat{a}_{n+g}^{\dag}\hat{a}_{l}\hat{a}_{n}$
in $H_{d}$, $\hat{a}_{l}\hat{a}_{n}$ destroys as much momentum as
$\hat{a}_{l-g}^{\dag}\hat{a}_{n+g}^{\dag}$ creates. This can be mathematically
expressed as that for each quartet $l,n,l-g,n+g\in J$, akin to the
calculation in Eq.~(\ref{eq:aaaa}), we have 
\begin{equation}
\left[\hat{a}_{l-g}^{\dag}\hat{a}_{n+g}^{\dag}\hat{a}_{l}\hat{a}_{n},\sum_{j\in J}k_{j}\hat{a}_{j}^{\dag}\hat{a}_{j}\right]=\left(-k_{l-g}-k_{n+g}+k_{l}+k_{n}\right)\hat{a}_{l-g}^{\dag}\hat{a}_{n+g}^{\dag}\hat{a}_{l}\hat{a}_{n}=0,
\end{equation}
which proves $\left[H_{d},P_{d}\right]=0.$

The discretized system $H_{d}$ is now structure-preserving and can,
in principle, be simulated on a quantum computer. However, in the
present study, travel along the other direction---we dequantize $H_{d}$
into a classical Hamiltonian function, 
\begin{align}
H_{d} & =H_{0}+H_{1},\,\,\label{eq:Hddequ}\\
H_{0} & =\sum_{l\in J}\frac{k_{l}^{2}\hbar^{2}}{2m}a_{l}^{*}a_{l},\thinspace\thinspace H_{1}=\sum_{l,n,l-g,n+g\in J}^{g\neq0}\frac{4\pi q^{2}}{L^{3}k_{g}^{2}}a_{l-g}^{*}a_{n+g}^{*}a_{l}a_{n}\thinspace,
\end{align}
where $\hat{a}_{l}^{\dagger}$ and $\hat{a}_{l}$ have been dequantized
into c-numbers $a_{l}^{*}$ and $a_{l}$, respectively. In the dequantized
theory, $H_{d}$ becomes the total energy, which splits into $H_{0}$,
the kinetic energy of dequantized particles, and $H_{1},$ the potential
energy between the dequantized particles. The equation of motion for
$a_{j}\,(j\in J)$ is

\begin{equation}
\dot{a}_{j}=\frac{1}{i\hbar}\left\{ a_{j},H_{d}\right\} =\frac{-k_{j}^{2}i\hbar}{2m}a_{j}+\frac{2\pi q^{2}}{i\hbar L^{3}}\sum_{l,l-g,j+g\in J}^{g\neq0}\frac{1}{k_{g}^{2}}a_{l}^{*}a_{l-g}a_{j+g}\thinspace.\label{eq:adotdequ}
\end{equation}
Because of the one-to-one correspondence between commutation relation
$\left[\,,\,\right]$ and the Poisson bracket $\left\{ \,,\,\right\} $,
the dequantized system conserves energy $H_{d}$, number of particles
$N_{d}=\sum_{l\in J}a_{l}^{*}a_{l}$, and total momentum $P_{d}=\sum_{l\in J}k_{l}a_{l}^{*}a_{l}$.

What physics does the finite-dimensional dequantized system (\ref{eq:Hddequ})
and (\ref{eq:adotdequ}) describe? It turns out it represents a structure-preserving
discretization of the Schrödinger-Poisson (SP) equations,

\begin{align}
\frac{\partial\psi}{\partial t} & =\frac{1}{i\hbar}\left[-\frac{\hbar^{2}}{2m}\nabla^{2}+q\phi\right]\psi\,,\label{eq:Schrodinger}\\
\nabla^{2}\phi & =-4\pi q\left(\psi^{*}\psi-\frac{N}{L^{3}}\right)\,.\label{eq:Poisson}
\end{align}
Here, $\psi(\boldsymbol{x},t)$ is the wave function of one particle.
To establish this fact, we work backward by finding a specific discretization
scheme for Eqs.\,(\ref{eq:Schrodinger}) and (\ref{eq:Poisson})
such that the resulting discretization agrees with Eqs.~(\ref{eq:Hddequ})
and (\ref{eq:adotdequ}).

To achieve this goal, we first reformulate Eqs.\,(\ref{eq:Schrodinger})
and (\ref{eq:Poisson}) as a Hamiltonian system defined by the Hamiltonian
functional 
\begin{align}
H(\psi,\psi^{*}) & =\int d^{3}\boldsymbol{x}\left\{ \psi^{*}\left[-\frac{\hbar^{2}}{2m}\nabla^{2}+\frac{1}{2}q\phi\right]\psi-\frac{N}{2}\frac{q}{L^{3}}\phi\right\} \,.\label{eq:H}
\end{align}
Here, it is understood that $\phi(\boldsymbol{x},t)$ is a function
determined by $\psi(\boldsymbol{x},t)$ via Eq.\,(\ref{eq:Poisson})
at every $t,$ since $\phi(\boldsymbol{x},t)$ cannot be cast as an
independent dynamical field. It can be shown that Eq.\,(\ref{eq:Schrodinger})
is equivalent to 
\begin{equation}
\frac{\partial\psi}{\partial t}=\frac{1}{i\hbar}\frac{\delta H}{\delta\psi^{*}}\,.
\end{equation}
Here, $\delta H/\delta\psi^{*}$ denotes functional derivative. Note
that $H$ depends on $\psi$ and $\psi^{*}$ explicitly and also implicitly
through $\phi(\boldsymbol{x},t).$ Both contributions need to be included
in the calculation of $\delta H/\delta\psi^{*}$.

We discretize the Hamiltonian in Eq.\,(\ref{eq:H}) in terms of Fourier
components. Let 
\begin{align}
\psi(\boldsymbol{x},t) & =\sum_{l}a_{l}(t)\exp(ik_{l}\cdot\boldsymbol{x})L^{-3/2}\,,\\
\phi(\boldsymbol{x},t) & =\sum_{l}\phi_{l}(t)\exp(ik_{l}\cdot\boldsymbol{x})L^{-3/2}\,,
\end{align}
From Eq.\,(\ref{eq:Poisson}), we have 
\begin{equation}
\phi_{l}=4\pi q\frac{1}{k_{l}^{2}}\sum_{n}a_{n}^{*}a_{l+n},\ (l\ne0).
\end{equation}
The $0$-th component of $\phi$ is an arbitrary constant, $\phi_{0}=\text{const.}\,.$
In terms of $a_{l}(t),$ we find 
\begin{equation}
H(\psi,\psi^{*})=\sum_{l}\frac{k_{l}^{2}\hbar^{2}}{2m}a_{l}^{*}a_{l}+\sum_{l,n,g\neq0}\frac{4\pi q^{2}}{L^{3}k_{g}^{2}}a_{l-g}^{*}a_{n+g}^{*}a_{l}a_{n}\thinspace,
\end{equation}
which reduces to $H_{d}$ in Eq.\,(\ref{eq:Hddequ}) once the truncation
condition $l,n,l-g,l+g\in J$ is imposed. Thus, we have proved that
the dequantized system (\ref{eq:Hddequ}) and (\ref{eq:adotdequ})
is a finite-dimensional, structure-preserving discretization of the
SP system.

The correspondence between VP and SP systems had been established
\citep{Husimi1940,Moyal1949,Feix1970,Cartwright1976,Bertrand1980,Kopp2017,Mocz2018,Cappelli2024}.
In its simplest form, the phase space distribution $f(\boldsymbol{x},\boldsymbol{v},t)$
in the VP system can be identified with the Wigner function constructed
from the wave function $\psi(\boldsymbol{x},t)$ in the SP system,
\begin{equation}
f(\boldsymbol{x},\boldsymbol{v},t)=\frac{1}{(2\pi\hbar/m)^{3}}\int_{\mathbb{R}^{3}}d^{3}\boldsymbol{y}\,\exp\!\Big(-\frac{i}{\hbar}m\boldsymbol{v}\cdot\boldsymbol{y}\Big)\,\psi^{*}\Big(\boldsymbol{x}+\frac{\boldsymbol{y}}{2}\Big)\,\psi\Big(\boldsymbol{x}-\frac{\boldsymbol{y}}{2}\Big).\label{Eq:Wigner}
\end{equation}
It is worth emphasizing that the distribution function $f(\boldsymbol{x},\boldsymbol{v},t)$
constructed this way is not semi-positive definite---a trade-off
for the simplicity of its form. The negative values of $f(\boldsymbol{x},\boldsymbol{v},t)$
arise from quantum effects \citep{Moyal1949,Feix1970,Hansen1973,Bertrand1980},
which are assumed to be small for the classical problems under study.
Husimi \citep{Husimi1940} showed that the Wigner function can be
smoothed with a minimum-uncertainty Gaussian kernel to produce a non-negative
phase space function now known as the Husimi function. Cartwright
\citep{Cartwright1976} extended this idea by proving that if the
Wigner function is smoothed with any kernel whose width is not smaller
than that of the minimum-uncertainty state, the resulting distribution
remains non-negative for all quantum states. Together, these works
establish that by appropriate coarse-graining, the Wigner transformation
can yield a non-negative quasi-probability distribution. With this
correspondence, Eqs.\,(\ref{eq:Hddequ}) and (\ref{eq:adotdequ})
furnish a finite-dimensional, structure-preserving discretization
of the VP system. This dequantized particle algorithm offers several
distinct advantages over conventional classical approaches. Most notably,
it evolves an equation system defined in 3D configuration space, whereas
classical algorithms operate in 6D phase space. This dimensional reduction
suggests that the dequantized particle algorithm can represent the
physical system's information more efficiently under suitable conditions.

Finally, we comment on the computational complexity of the dequantized
particle algorithm. For a system with $M$ dequantized particles,
the complexity of one-step time-advance map is ostensibly $M^{3},$
judging from summation of the $a_{l}^{*}a_{l-g}a_{j+g}$ terms in
Eq.\,(\ref{eq:adotdequ}). However, this summation consists of convolution
summations, which can be efficiently calculated in $M\log M$ operations,
akin to the standard spectrum and PIC algorithms.

We now present a numerical example of the dequantized particle algorithm
using the test problem of 1D two-stream instability. For numerical
calculations, we normalize length by $L,$ time by $\omega_{p}$,
and energy by $mL^{2}\omega_{p}^{2}$. Here, $\omega_{p}\equiv\sqrt{4\pi Nq^{2}/L^{3}m}$
is the plasma frequency. Specifically, the normalized variables are
\begin{equation}
\bar{\boldsymbol{x}}=\boldsymbol{x}/L,\,\,\,\bar{t}=t\omega_{p},\thinspace\thinspace\thinspace\bar{\psi}=\psi L^{3/2}N^{-1/2},\,\,\,\bar{a}_{j}=a_{j}N^{-1/2},\,\,\,\bar{\phi}=\frac{q\phi}{mL^{2}\omega_{p}^{2}},\,\,\,\bar{H}=\frac{H}{\omega_{p}}.
\end{equation}
In normalized variables, Eq.\,(\ref{eq:adotdequ}) is 
\begin{equation}
\dot{a}_{j}=\frac{\delta}{2i}a_{j}k_{j}^{2}+\frac{1}{i\delta}\sum_{l,l-g,j+g\in J}^{g\neq0}\frac{1}{k_{g}^{2}}a_{l}^{*}a_{l-g}a_{j+g}\thinspace,\,\,\,j\in J\,.
\end{equation}
Here, the over bar ``$\bar{\thinspace\thinspace\,}$'' for normalized
variables has been dropped for easy notation and 
\begin{equation}
\delta\equiv\frac{\hbar}{L^{2}\omega_{p}m}=\left(\frac{\hbar}{Lmv_{th}}\right)\left(\frac{\lambda_{D}}{L}\right)
\end{equation}
is a dimensionless parameter measuring the quantum effect using the
characteristics of a classical plasma. Here, $v_{th}$ is the thermal
velocity of the plasma and $\lambda_{D}$ is the Debye length. For
classical dynamics, $\delta\ll1.$

\begin{figure}
\includegraphics[width=5cm]{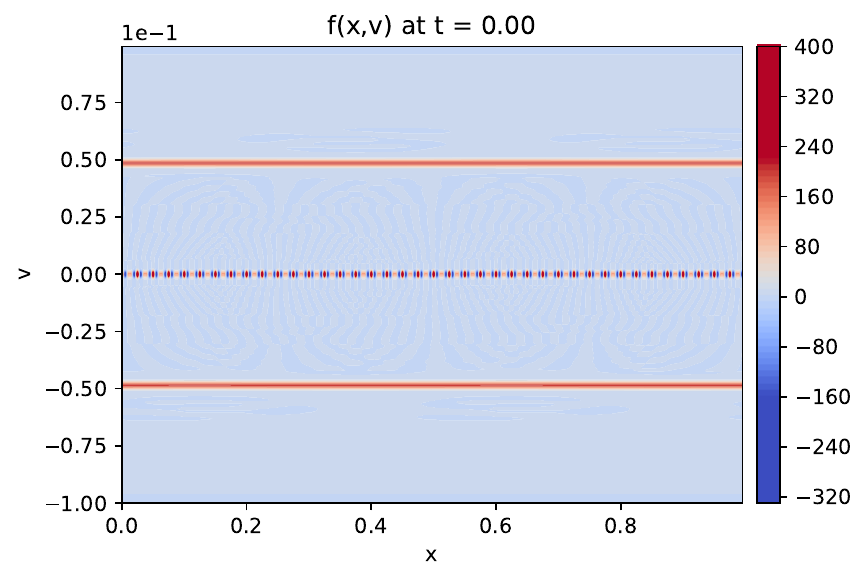}\includegraphics[width=5cm]{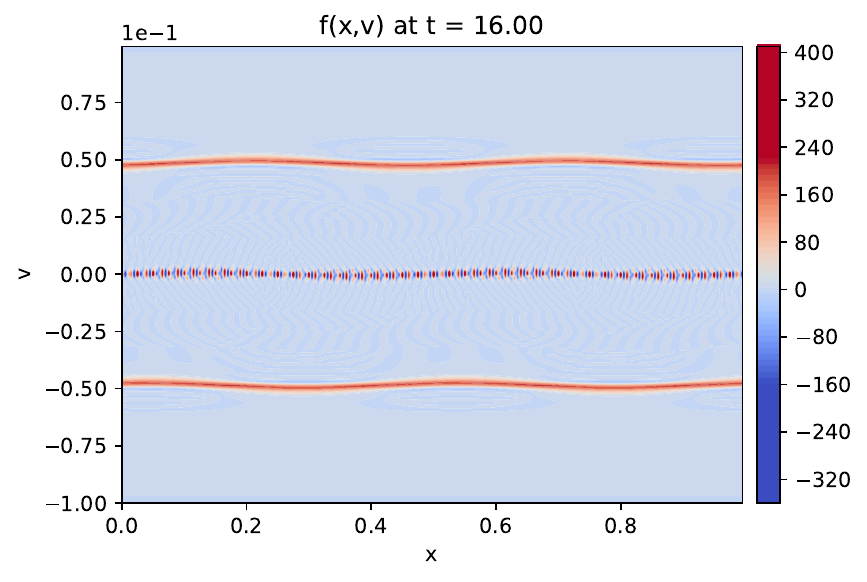}\includegraphics[width=5cm]{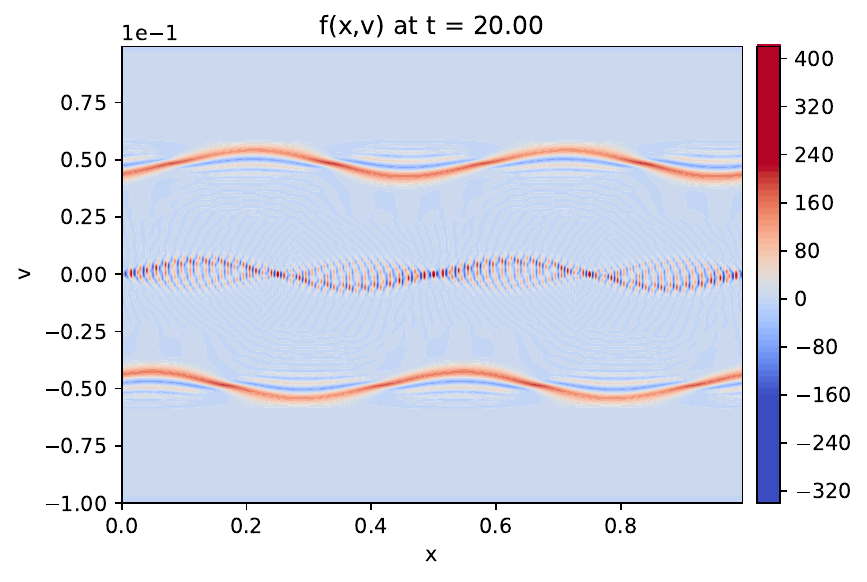}

\includegraphics[width=5cm]{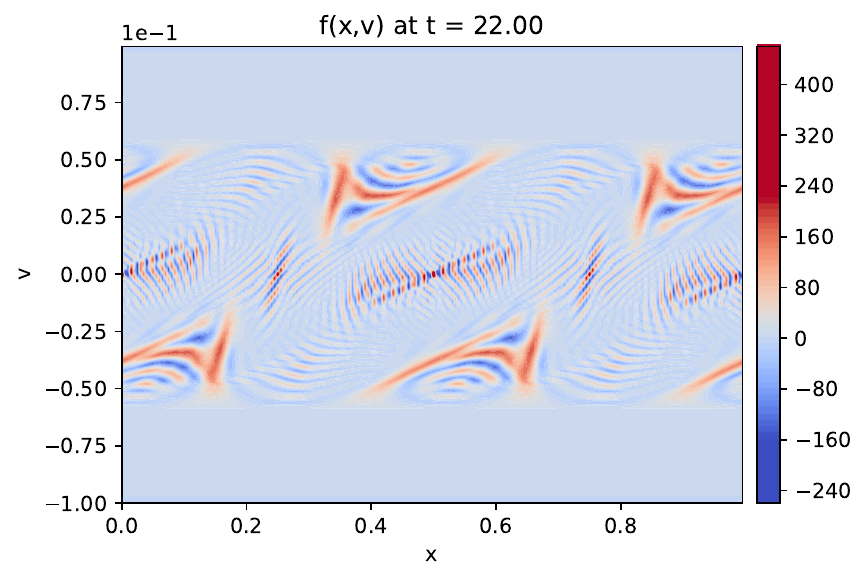}\includegraphics[width=5cm]{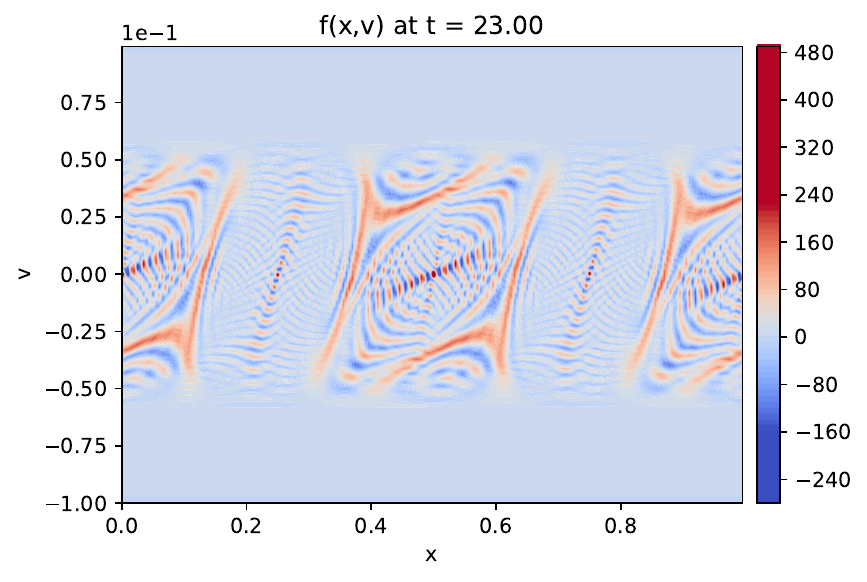}\includegraphics[width=5cm]{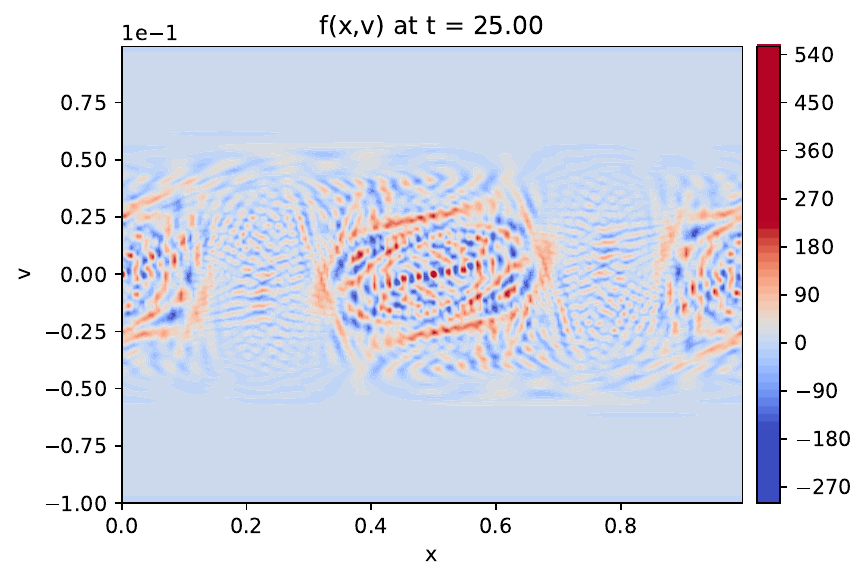}

\caption{Nonlinear two-stream instability simulated with 97 dequantized particles.
Shown is the distribution function $f(x,v,t)$ at $t=0,16,20,22,23,25$,
constructed using Eq.\,(\ref{Eq:Wigner}) from $a_{j}\,(-48\le j\le48).$
Phase-space vortex structures develop during the instability. \label{fig:f}}
\end{figure}

\begin{figure}
\includegraphics[width=5cm]{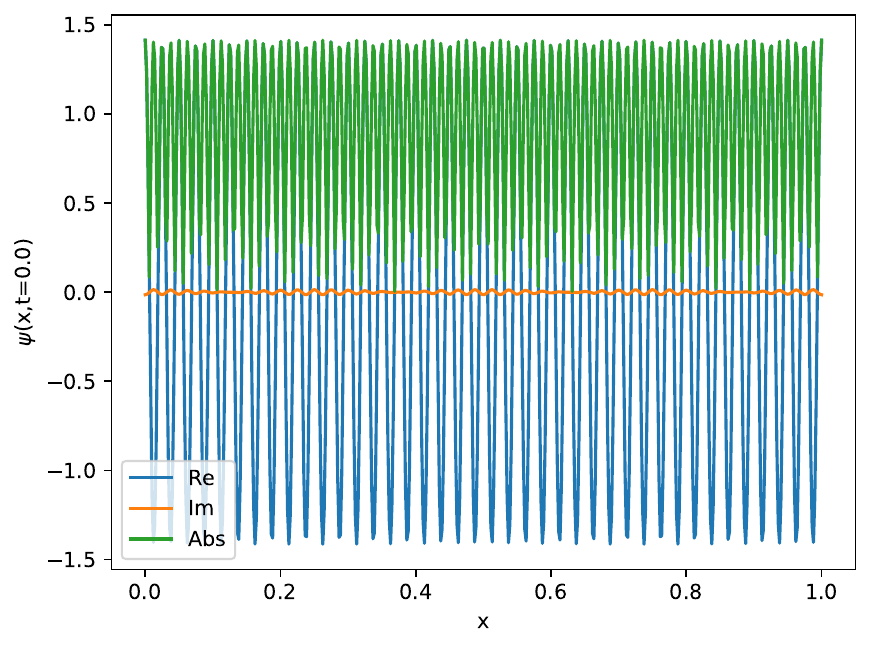}\includegraphics[width=5cm]{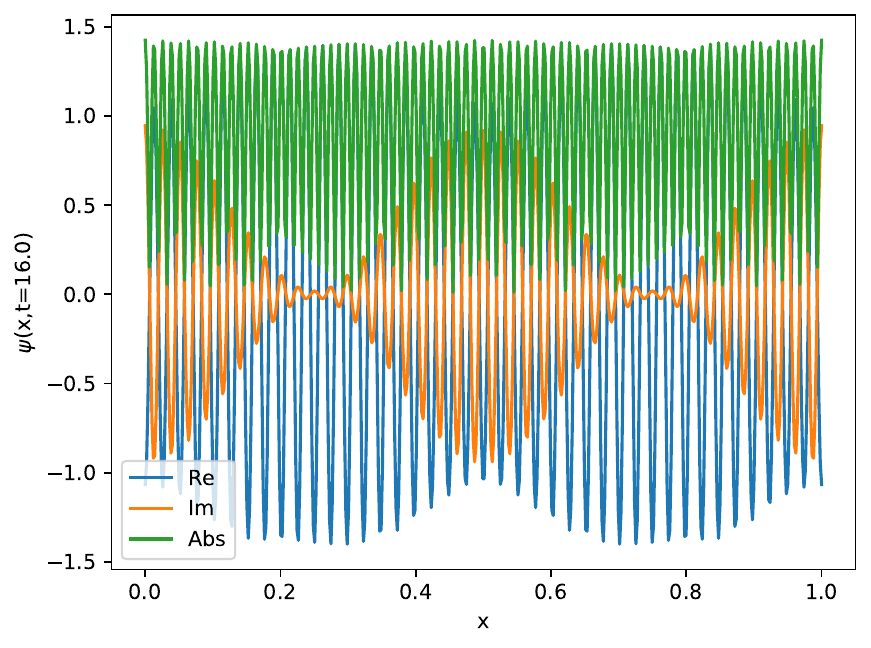}\includegraphics[width=5cm]{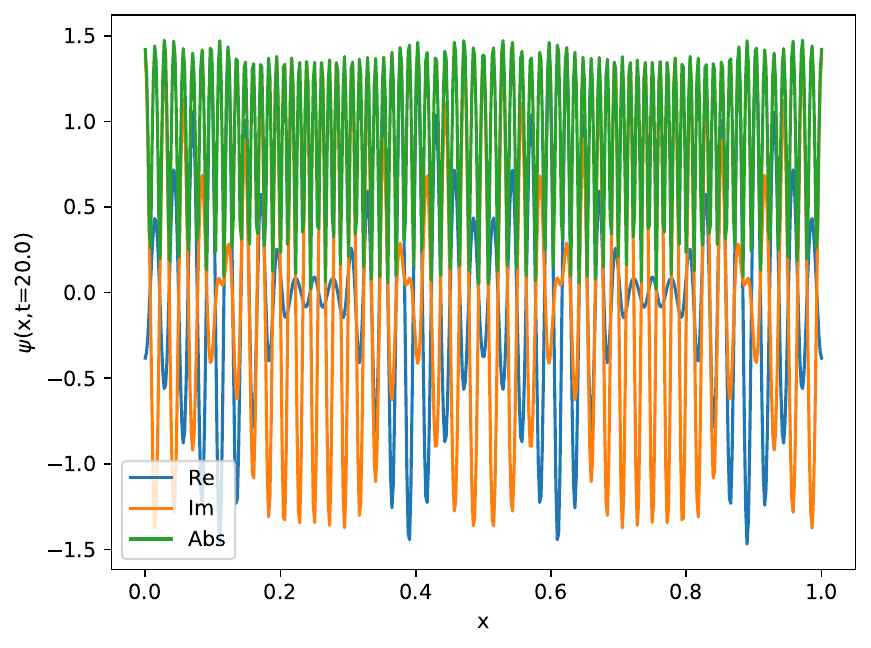}

\includegraphics[width=5cm]{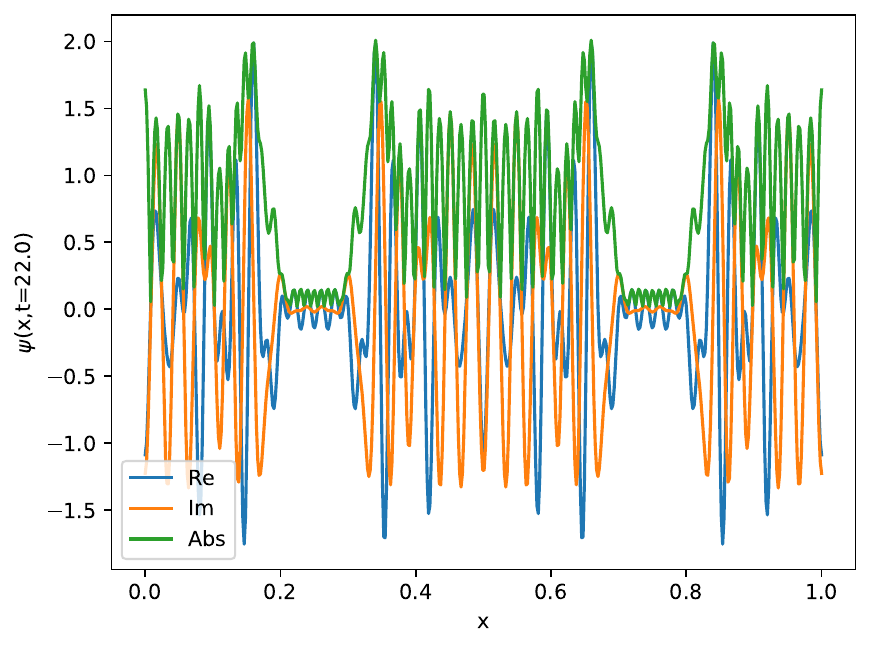}\includegraphics[width=5cm]{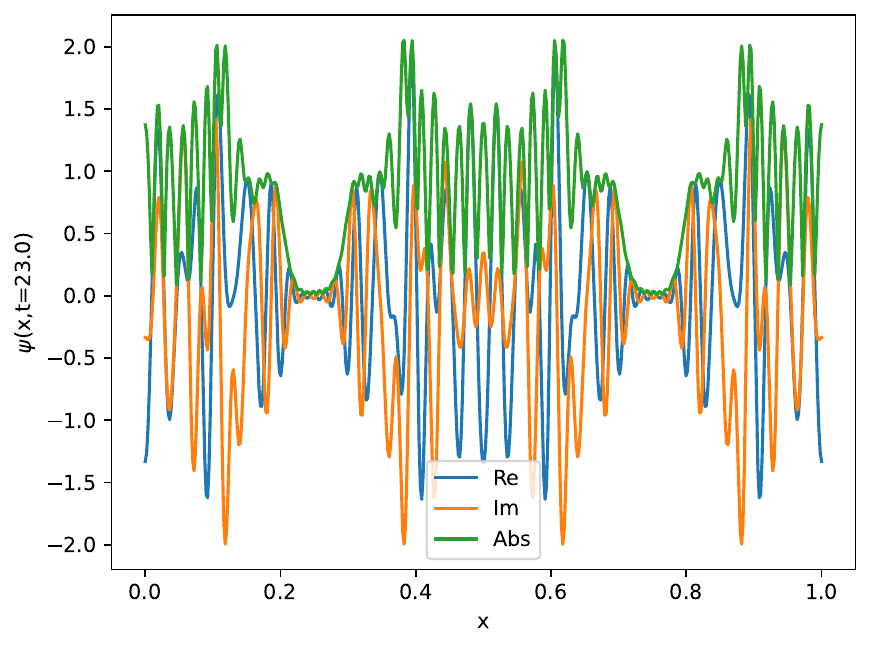}\includegraphics[width=5cm]{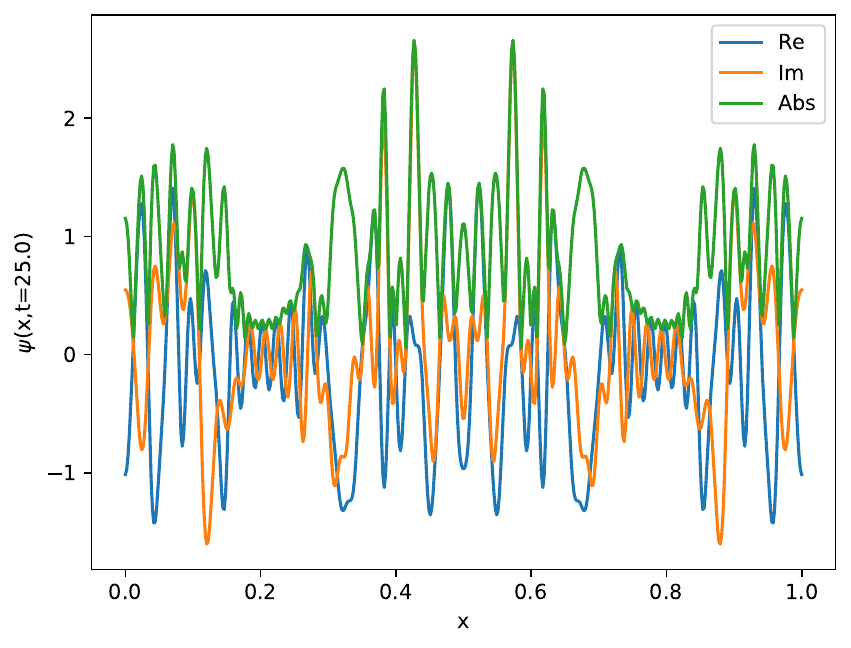}

\caption{\label{fig:psi}The wave function $\psi(x,t)$ at $t=0,16,20,22,23$,
and 25 is constructed from the 97 dequantized particles $a_{j}\,(-48\le j\le48).$
In the linear phase, the wave function is dominated by a small number
of dequantized particles, whereas in the nonlinear phase, more dequantized
particles participate in the dynamics. }
\end{figure}

\begin{figure}
\includegraphics[width=5cm]{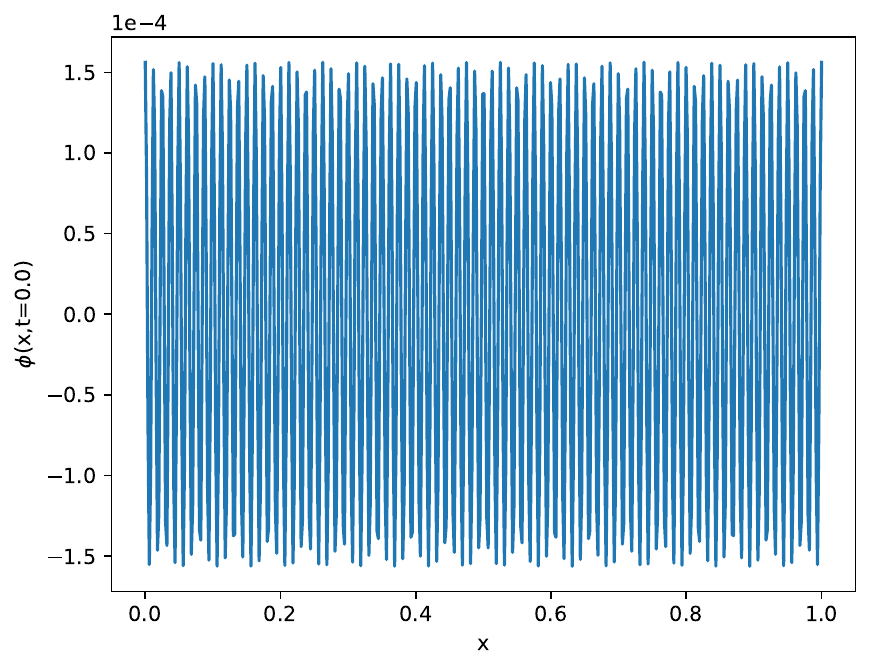}\includegraphics[width=5cm]{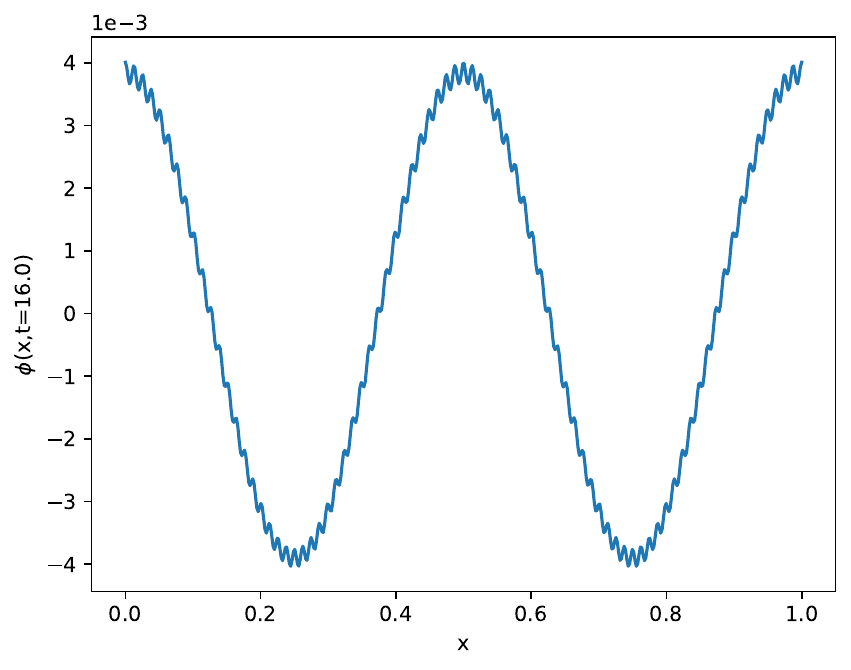}\includegraphics[width=5cm]{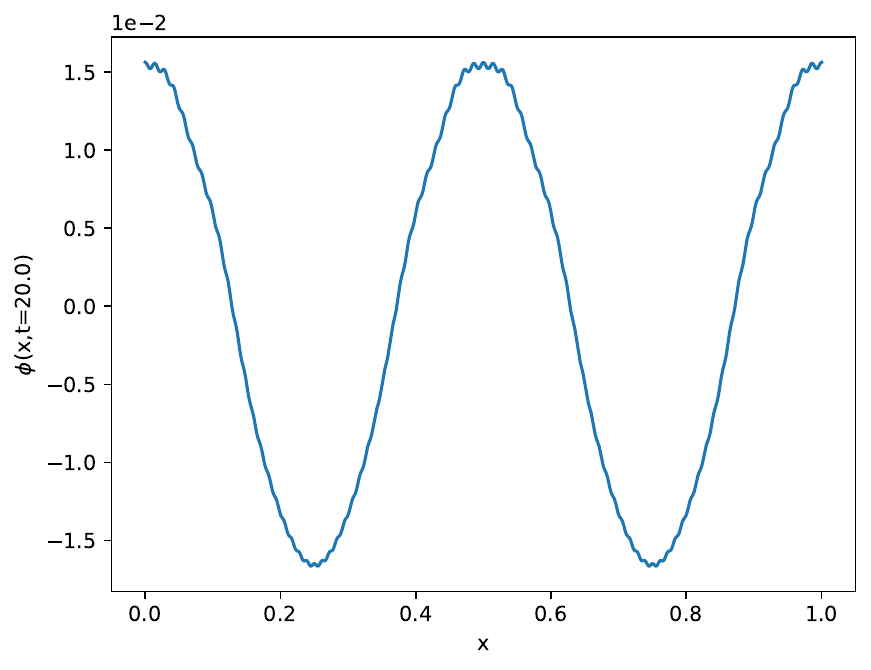}

\includegraphics[width=5cm]{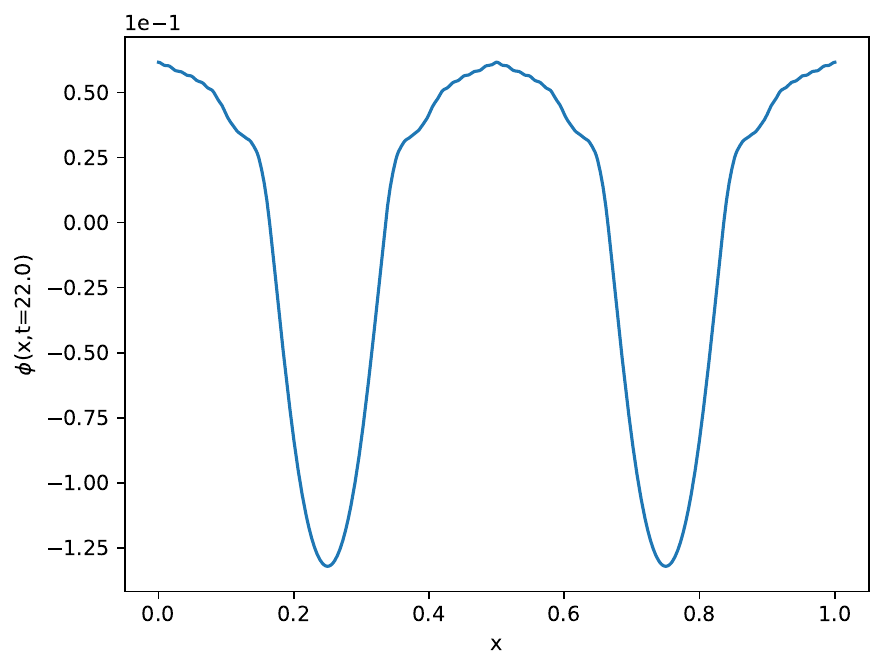}\includegraphics[width=5cm]{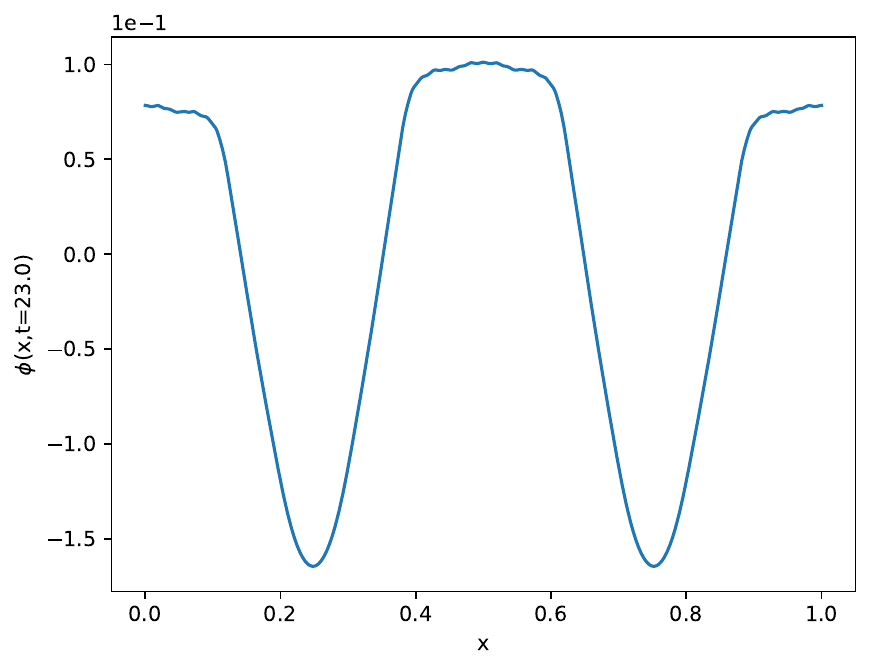}\includegraphics[width=5cm]{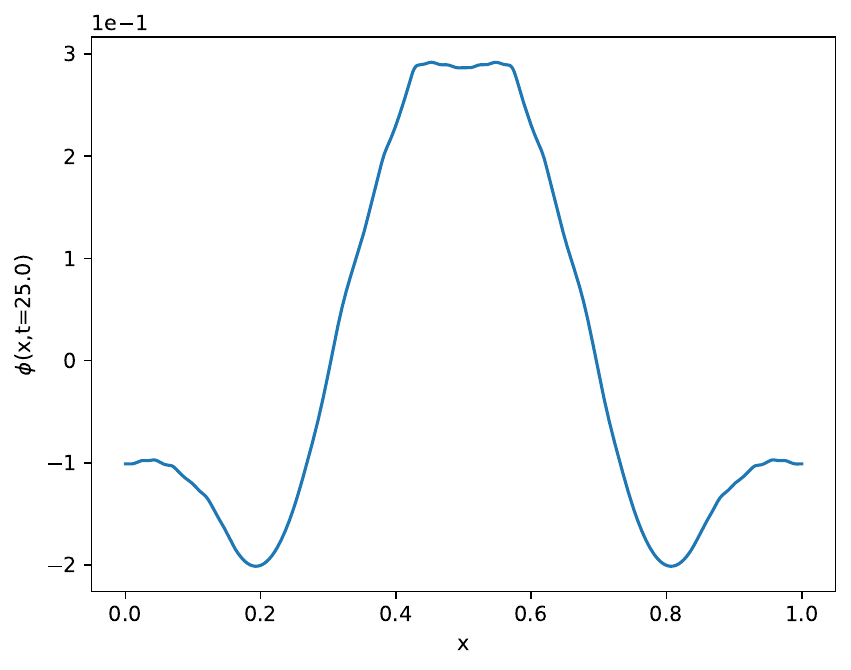}

\caption{\label{fig:phi}The photon field $\phi(x,t)$ at $t=0,16,20,22,23,$
and $25$ is constructed from the 97 dequantized particles $a_{j}\,(-48\le j\le48).$
In the linear phase, the potential energy $H_{1}$ increases monotonically,
while in the nonlinear phase, the mediating photon field deviates
significantly from being monochromatic. To illustrate the structure
of the field during the instability, different scales are used for
the different panels.}
\end{figure}

\begin{figure}
\includegraphics[width=6cm]{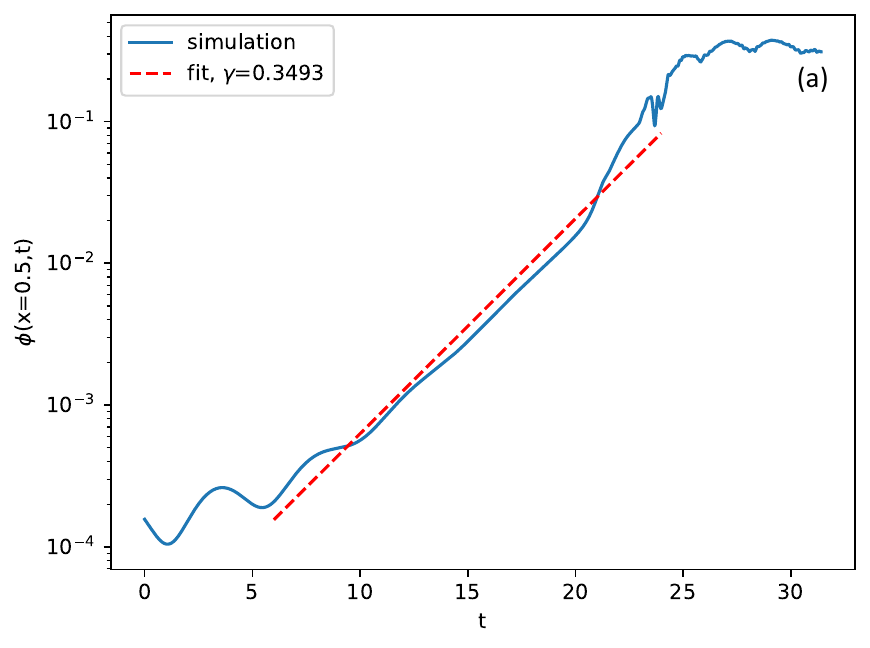}\includegraphics[width=6cm]{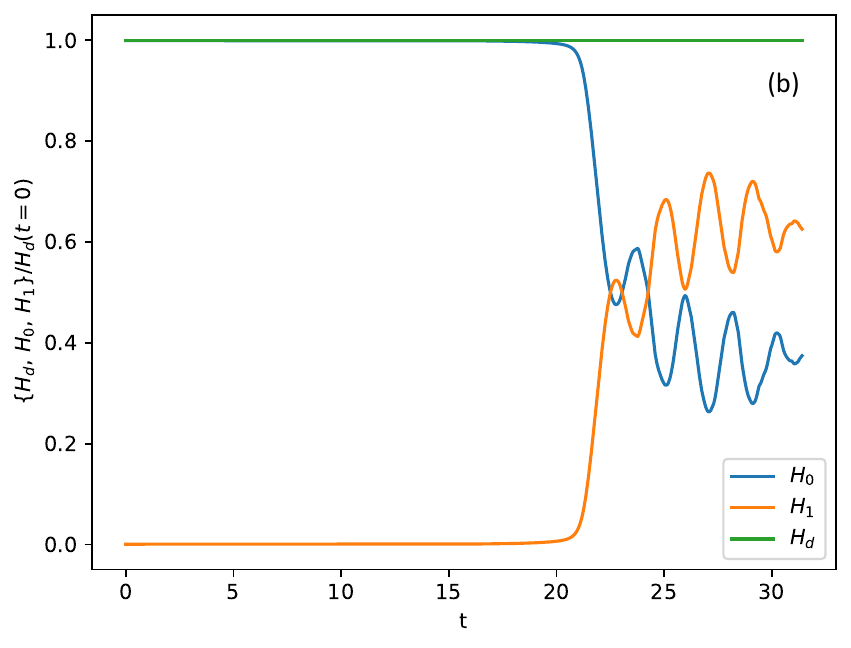}

\includegraphics[width=6cm]{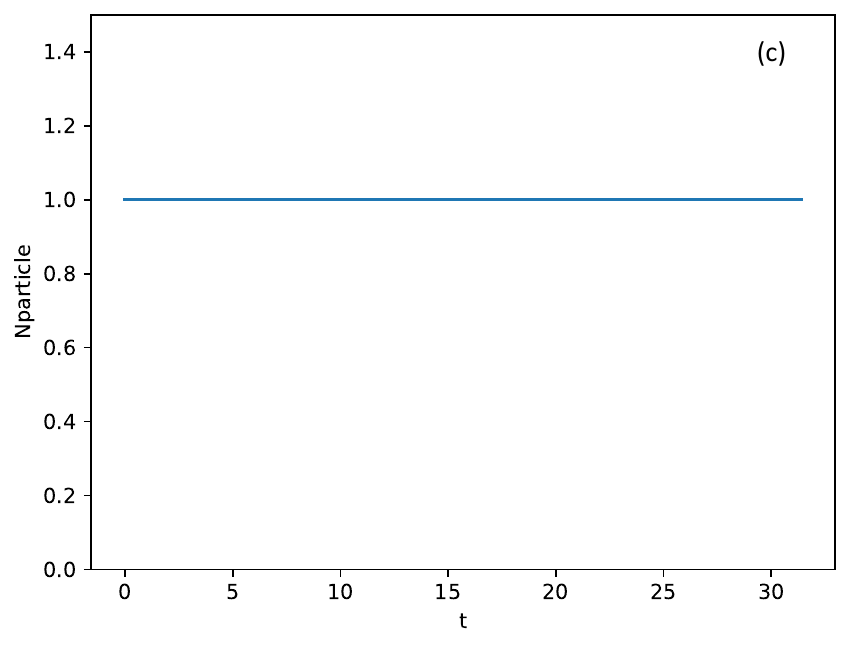}\includegraphics[width=6cm]{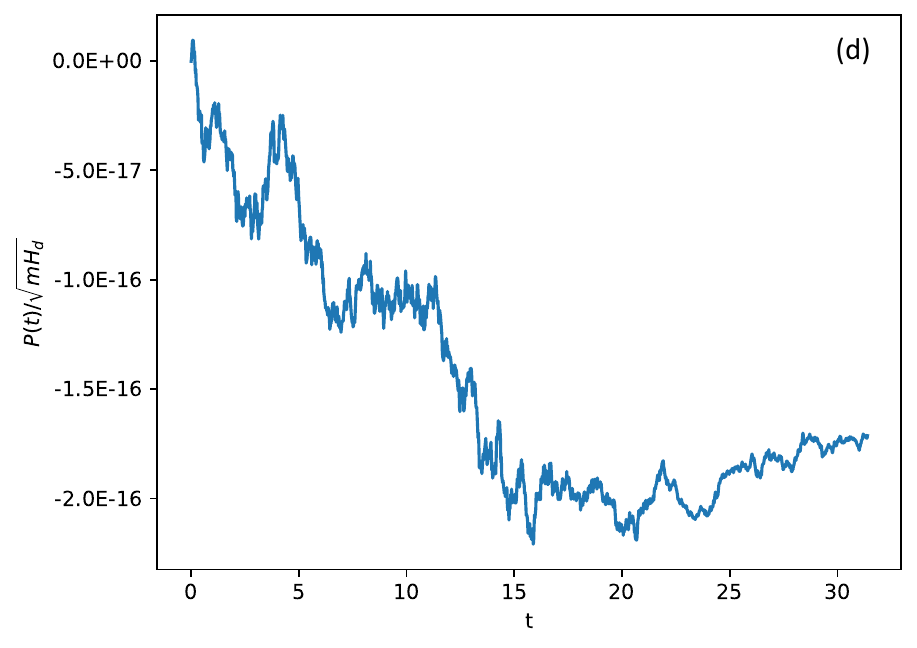}

\caption{\label{fig:gamma}Nonlinear two-stream instability simulated with
97 dequantized particles. (a) Growth rate; (b) time history of energy;
(c) number of particles; (d) momentum. The initial momentum is zero
for the two-stream distribution, and the momentum variation is measured
by $\sqrt{mH_{d}}$. Energy, number of particles, and momentum are
well conserved to high precision.}
\end{figure}

To load in an initial two-stream distribution with peaked population
at $v=\pm V_{0},$ we follow Ref.~\citep{Bertrand1980} to let
\begin{equation}
\psi(x,t=0)=\sqrt{2}\cos\left(\frac{xV_{0}}{\delta}\right)\exp\left(-\frac{i\epsilon V_{0}}{\delta k}\cos(kx)\right),
\end{equation}
where an initial perturbation at wavelength $k$ and amplitude $\epsilon$
has also been included. Displayed in Figs.\,\ref{Fig:FD}-\ref{fig:gamma}
are simulation results for the case of $(V_{0},\delta,k,\epsilon)=(0.04854,1.9\times10^{-4},4\pi,5\times10^{-4})$
with 97 dequantized particles, i.e., $J=\left\{ n\in\mathbb{N}|-48\le n\le48\right\} .$
The simulation was run from $t=0$ to $30$. The time history of $f(x,v,t)$,
constructed using Eq.\,(\ref{Eq:Wigner}), is displayed in Fig.\,\ref{fig:f},
which illustrates the development of phase-space vortex structures
during the instability. As noted previously, the negative values of
$f(x,v,t)$ are associated with the quantum effects \citep{Moyal1949,Feix1970,Hansen1973,Bertrand1980},
which manifest as small scale structures with wavelengths proportional
to $\delta$ in this problem. If preferred, the non-negative Husimi
function \citep{Husimi1940,Cartwright1976} can be used as an alternative.
By The time history of $\psi(x,t)$ and $\phi(x,t)$ are shown in
Figs.\,\ref{fig:psi} and \ref{fig:phi}, respectively. The two-stream
instability grows exponentially in the linear phase and saturates
approximately at $t=25.$ The growth rate at the linear phase is measured
to be $\gamma=0.3493$ {[}Fig.\,\ref{fig:gamma}(a){]}, which agrees
well with the theoretical growth rate $\gamma_{\text{theory}}=0.3536,$
solved for from the analytical dispersion relation 
\begin{equation}
\frac{1}{2(\omega+kV_{0})^{2}}+\frac{1}{2(\omega-kV_{0})^{2}}=1.
\end{equation}

The time history of the total energy, number of particles, and total
momentum are plotted in Fig.\,\ref{fig:gamma}. All three quantities
are well conserved to high precision. In the linear phase, the wave
function is dominated by a small number of dequantized particles,
the kinetic energy $H_{0}$ decreases monotonically, the potential
energy $H_{1}$ increases monotonically, whereas the total energy
$H_{d}$ is conserved. In the nonlinear phase, more dequantized particles
participate in the dynamics, the mediating photon field deviates significantly
from being monochromatic, and both $H_{0}$ and $H_{1}$ oscillate
at approximately the same level.

In conclusion, we have developed a dequantization algorithm for the
VP system, termed the dequantized particle algorithm, by dequantizing
the many-body quantum Hamiltonian in its second-quantized form. The
resulting finite-dimensional system preserves the key conservation
laws and furnishes a structure-preserving discretization of the SP
equations, which connects naturally to the VP system through the Wigner
transformation. This formulation enables an alternative simulation
strategy in 3D configuration space rather than the full 6D phase space
for the VP system. The example of the two-stream instability using
97 dequantized particles, which correspond to the Fourier modes of
the VP system, confirms that the algorithm accurately captures the
expected physical behavior and conserves invariants to high precision.
These results highlight the potential of the dequantization approach
to provide new pathways for developing efficient, structure-preserving
classical algorithms as well as quantum algorithms for plasma kinetic
simulations \citep{shi2018,ShiThesis,Joseph2020,Shi2021,Shi2021a,Dodin2021,Novikau2022,May2023,May2024,May2024a}. 
\begin{acknowledgments}
This research is supported by the U.S. Department of Energy (DE-AC02-09CH11466).
J. M. Molina acknowledges support from the National Science Foundation
Graduate Research Fellowship under Grant No. KB0013612. 
\end{acknowledgments}

\bibliographystyle{apsrev4-2}
\bibliography{DA}

\end{document}